\documentclass{article}
\usepackage{amsfonts,amssymb,latexsym,amsmath,amscd}
\usepackage{times,mathptm,color}
\usepackage{anysize}
\marginsize{1in}{1in}{1in}{1in}
\usepackage{graphicx}

    {
    \smallskip
    \refstepcounter{theorem}
    \noindent
    {\bf Example \arabic{section}.\arabic{theorem}} \ \ }
    {\hspace*{\fill}{$\Diamond$}
    \smallskip}

    {
    \smallskip
    \refstepcounter{theorem}
    \noindent
    {\bf Remark \arabic{section}.\arabic{theorem}} \ \ }
    {\hspace*{\fill}{$\Diamond$}
    \smallskip}

\newenvironment{appremark}
    {
    \smallskip
    \refstepcounter{theorem}
    \noindent
    {\bf Remark \Alph{section}.\arabic{theorem}} \ \ }
    {\hspace*{\fill}{$\Diamond$}
    \smallskip}

\newenvironment{definition}
    {
    \smallskip
    \refstepcounter{theorem}
    \noindent
    {\bf Definition \arabic{section}.\arabic{theorem}} \ \ }
    {\hspace*{\fill}{\ }
    \smallskip}

\newenvironment{appdefinition}
    {
    \smallskip
    \refstepcounter{theorem}
    \noindent
    {\bf Definition \Alph{section}.\arabic{theorem}} \ \ }
    {\hspace*{\fill}{\ }
    \smallskip}

\newenvironment{proof}[1][]
    {
    \noindent
    {\bf Proof{#1}:  }
    }
    {\hspace*{\fill}{$\Box$}\smallskip}

    {
    \noindent
    {\bf Sketch{#1}:  }
    }
    {\hspace*{\fill}{$\Box$}\smallskip}

    {
    \noindent
    {\bf Reference:  }
    }
    {\hspace*{\fill}{$\odot$}\smallskip}

\newtheorem{theorem}{Theorem}[section]
\newtheorem{proposition}[theorem]{Proposition}
\newtheorem{lemma}[theorem]{Lemma}
\include{Qcircuit}

\newcommand{\controlu}{*-=[][F]{\phantom{\bullet}}}

\begin{document}

\title{Asymptotically Optimal Quantum Circuits for $d$-level Systems}

\author{
\begin{tabular}{ccccc}
Stephen S. Bullock$^{1}$
& \;\;\;\; & 
Dianne P. O'Leary$^{1,3}$ 
& \;\;\;\;
& Gavin K. Brennen$^2$ \\
\footnotesize \tt stephen.bullock@nist.gov & &
\footnotesize \tt oleary@cs.umd.edu & &
\footnotesize \tt gavin.brennen@nist.gov \\
\end{tabular}
}

\date{\today}

\maketitle

\begin{center}
\begin{tabular}{l}
$^1$ National Institute of Standards and Technology, \\
\quad Mathematical and Computational Sciences Division, 
Gaithersburg, Maryland 20899-8910 \\
$^2$ National Institute of Standards and Technology, Atomic Physics Division,
Gaithersburg, Maryland, 20899-8420 \\
$^3$ University of Maryland, Department of Computer Science, College Park,
Maryland 20742. \\
$\; \; \;${The work of this author was supported in part
by the National Science Foundation under Grant CCR-0204084.} 
\end{tabular}
\end{center}

\begin{abstract}
As a qubit is a two-level quantum system whose state space is spanned
by $\ket{0}$, $\ket{1}$, so a qudit is a $d$-level quantum system whose
state space is spanned by $\ket{0},\cdots,\ket{d-1}$.  
Quantum computation has stimulated much recent
interest in algorithms factoring unitary evolutions of an $n$-qubit state
space into component two-particle unitary evolutions.  In the absence of
symmetry, Shende, Markov and Bullock use Sard's theorem to prove that
at least $C 4^n$ two-qubit
unitary evolutions are required, while Vartiainen, M{\"o}tt{\"o}nen,
and Salomaa ({\tt VMS})
use the $QR$ matrix factorization and Gray codes
in an optimal order construction involving two-particle evolutions.  
In this work, we note that
Sard's theorem demands $C d^{2n}$ two-qudit unitary
evolutions to construct a generic (symmetry-less) $n$-qudit evolution.
However, the {\tt VMS} result applied to virtual-qubits only recovers
optimal order in the case that $d$ is a power of two.  
We further construct a $QR$ decomposition for
$d$-multi-level quantum logics, proving a sharp asymptotic of
$\Theta(d^{2n})$ two-qudit gates
and thus closing the complexity question for
all $d$-level systems ($d$ finite.)  Gray codes are not required,
and the optimal $\Theta(d^{2n})$ asymptotic also applies to gate
libraries where two-qudit interactions are restricted by a choice
of certain architectures.
\end{abstract}

\noindent
\begin{center}
\hfill{\space}
{\bf AMS(MOS) subj:} 81P68, 
65F25 
\end{center}

\vbox{
\tableofcontents
}

\section{Introduction}

The dominant theoretical model of quantum computation is the quantum
circuit \cite{Deutsch:89} acting on quantum bits, or 
qubits \cite{Feynman:86}.  A qubit is a two-level quantum system,
whose complex Hilbert state space is spanned by kets $\ket{0}$ and
$\ket{1}$.  The labels within the ket are evocative of classical computer
logic.  
Yet using qubits, these logical values may be placed
in superposition, and moreover multiple qubits may be entangled.  A
\emph{qudit} \cite{Gottesman:99,Blume-KahoutEtAl:02}
is a generalization to quantum computing of a classical
multi-level logic \cite{Lawler:64}.  We fix $d \geq 2$ throughout and
consider the one-qudit state space
\begin{equation}
\label{eq:factor}
\mathcal{H}(1,d) \ = \ \mathbb{C} \ket{0} \oplus \mathbb{C} \ket{1} \oplus
\cdots \oplus \mathbb{C} \ket{d-1}.
\end{equation}
The decomposition is taken to be Hermitian orthonormal, and the $n$-qudit
state space then becomes $\mathcal{H}(n,d) = \otimes_1^n \mathcal{H}(1,d)=
\oplus_{c_1c_2\ldots c_n} \mathbb{C} \ket{c_1c_2\ldots c_n}$ with
$c_1c_2 \ldots c_n$ varying over all length $n$ integers in base $d$.

A quantum computation is a procedure
that takes a classical input string encoded in a quantum data state, processes this state using operations allowed by the laws of quantum mechanics, and finally measures the state to produce a classical output string.  The quantum processing can be realized 
as a unitary evolution on the state space.  The $\it{exact}$ universality theorem for quantum computation with qudits \cite{Brylinski:02} states that any unitary evolution on many qudits can be constructed to infinite precision using a finite sequence of single qudit and two-qudit unitaries or gates.  Any such sequence of quantum gates that transforms classical input to classical output is known as a quantum algorithm.   Of course, not all quantum algorithms are efficient.  As most functions
on bit-strings require exponentially many {\tt AND-OR-NOT} gates, so too
most unitary evolutions may only be realized with exponentially many
quantum gates.  Efficient quantum algorithms are usually defined as those using a number of single and two qudit gates whose size (complexity) is asymptotically bounded above by a polynomial in the number of qudits.

We say a function $h(n) \in \Omega[f(n)]$ if there is a constant $C$ 
so that $h(n)$ is at least
$Cf(n)$ for $n \geq 1$, and similarly $h(n) \in O[f(n)]$
if there is a second $C$ so that $h(n)$ is at most $C f(n)$.
We say that $h(n) \in \Theta[f(n)]$ when both hold.
Several choices of gate libraries are used in quantum
algorithms, but most admit asymptotically equivalent gate counts.  We
concentrate on two-qudit gates \cite{DiVincenzo:95}.
Thus, the complexity of a unitary evolution $U$
is that number $\ell$ for which we have a minimum length expression
\begin{equation}
U=U^1_{j_1 k_1} U^2_{j_2k_2}\cdots U^\ell_{j_\ell k_\ell}
\end{equation}
with each $U^p_{jk}$ a two-qudit 
($d^2 \times d^2$) operator acting exclusively on qudits $j$,$k$.  
An efficient computation should produce a family $\{ U_n\}_{n=1}^\infty$
of unitary operators
whose gate counts $\ell_{U_n}=h(n)$ satisfy $h(n) \in O(n^p)$, some $p > 0$.
As an example of this formalism, 
put $\omega= \mbox{e}^{2 \pi i/2^n}$
and consider the $n$-qubit Fourier transform 
$\mathcal{F}_n = \frac{1}{\sqrt{2^n}}
\sum_{j,k=0}^{2^n-1} \omega^{jk} \ket{k}\bra{j}$.
Known circuits for $\mathcal{F}_n$ require $O(n^2)$ gates 
\cite{NielsenChuang:00}, so that the Fourier
transform is an efficient quantum computation.
The extension to qudits is likewise an efficient quantum computation
\cite{Hoyer:97}.

It is typical to draw the factorization of $U$ in
Equation \ref{eq:factor} as a quantum
circuit, representing each qudit with a line and drawing a
gate connecting qudits $j$, $k$ for each $U_{jk}^q$.
Physical implementation of symmetry-less evolutions is
not practical when the number of qudits
$n$ is large, since the number of gates  required scales 
exponentially in $n$.  Yet circuits for generic unitaries are still
of interest.  First, they may improve subblocks of larger circuits
through a process of peephole optimization:  when many
consecutive two-qudit gates act on a small collection of qudits, we
compute the associated unitary evolution and substitute a
circuit of the sort presented here in hopes of decreasing the total
number of required operations.  Second, they are useful 
in translating circuits from gate libraries that include three 
and multi-qudit gates to two-qudit gates
when a physical system only conveniently allows for pairwise
interactions.
They may also be
used to translate an arbitrary gate library into a fault-tolerant 
library of qudit gates \cite{Knill:96}.  Finally, we note that
the symmetries that allow for polynomial-size quantum circuits are not 
well-understood.  Thus, producing efficient symmetry-less circuits
may provide insights into general design principles that might
also be useful in constructing or optimizing computations.

For qubits, Shende, Markov and Bullock have 
shown that $\Omega(4^n)$ two-qubit gates
are required, while a recent Letter
\cite{Vartiainen:04} provided a $O(4^n)$ construction.  
Thus we have  
a sharp asymptotic for symmetry-less $n$-qubit unitary evolution:
$\Theta(4^n)$ two-qubit gates are required.  The result does
\emph{not} readily extend to qudits, even though qudit systems may be
employed to emulate qubit systems and conversely.  
The lower bound generalizes to $C d^{2n}$ gates, but
na{\"i}ve emulations of the VMS circuit require 
asymptotically more gates than this.
Indeed, the best prior constructive upper bound
is $O(n^2 d^{2n})$ two-qudit gates \cite{MuthukrishnanStroud:00}.

The main result of our work is a constructive proof
that $\Theta(d^{2n})$ two-qudit gates are required
to implement an arbitrary $n$-qudit evolution without symmetry.
En route, we also prove that $\Theta(d^n)$ two-qudit gates suffice
for $n$-qudit state-synthesis.
The algorithm that produces the quantum circuit
is a variant of the $QR$ matrix-decomposition, cf. 
\cite{BarencoEtAl:95,Cybenko:01,Vartiainen:04}. 
Unlike an earlier qubit construction \cite{Vartiainen:04}, it
does not rely on a Gray code, either in base-two or base $d$.

The paper is organized as follows.
First in \S \ref{sec:lower},
we review the justification of the lower bound 
of $\Omega(d^{2n})$ and then
in \S \ref{sec:emulate}
we discuss the inadequacy of qubit emulation of qudits.
The remainder of the manuscript
describes an algorithm for constructing a 
circuit involving two-qudit gates, carefully 
showing that the number of gates is $O(d^{2n})$.
In \S \ref{sec:uniform_control}, we define a controlled
single qudit gate which applies a single qudit 
unitary conditioned on the state of multiple control qudits.  In particular, we show that
a $k$-controlled one-qudit unitary may be implemented in $O(k)$
two-qudit gates, given sufficient ancilla (helper) qudits.
In \S \ref{sec:Householder}, we describe our state-synthesis algorithm
and adapt it to a virtual Householder reflection
using singly controlled one-qudit operators.
In \S \ref{sec:upper}, we present our qudit-native quantum circuit
synthesis algorithm, and establish in \S \ref{sec:counts}
that it produces a universal circuit with at most
$O(d^{2n})$ two-qudit operations.

\section{The Lower Bound}
\label{sec:lower}

The lower bound argument
is similar to other lower bound arguments
\cite{ShendeMarkovBullock:03,BullockMarkov:04} 
in quantum computing using Sard's theorem from smooth topology.  The
theorem (loosely) states that almost no values of a smooth function are
critical values.  A well-known corollary (e.g. \cite{BullockMarkov:04})
then demands that for a smooth
map $f:M \rightarrow N$ that carries an $m$-dimensional manifold $M$
into an $n$-dimensional manifold $N$ for $m<n$, the set
$\mbox{image}(f)$ must be a measure zero subset of $N$.

We first set some notation.  By default, upper case letters indicate
either matrices or unitary operators.  We use $I_\ell$ to denote
an $\ell \times \ell$ identity matrix, and $A^\dagger = \overline{A}^T$ is
the adjoint of $A$.
Recall also
the Lie theory notation $U(q)=\{U \in \mathbb{C}^{q \times q} \; ; \;
U U^\dagger = I_q\}$.
Suppose then that we consider an expression associated to a fixed
circuit topology for two-qudit gates.  Namely, suppose we factor a
$U \in U(d^{2n})$ as in Equation \ref{eq:factor}.
Suppose moreover that we take $\ell$ and the tuples $(j_q, k_q)$ for 
$1 \leq q \leq \ell$ to be fixed.  Then by varying the $U_{j_q k_q}^q$
in $U(d^2)$, we obtain a map of smooth manifolds
$f: [U(d^2)]^\ell \rightarrow U(d^{n})$.  Now generically,
$\mbox{dim}_{\mathbb{R}}[U(q)]=q^2$.  Hence the smooth function implicit in the
circuit diagram of Equation \ref{eq:factor} carries a manifold of dimension
$\ell d^{4}$ into a manifold of dimension $d^{2n}$.  In order
for the image to not be measure $0$ for a fixed circuit diagram,
we require $\ell \geq d^{2n-4}$.  As there are only finitely many circuit
topologies holding fewer than $d^{2n-4}$ 
factors per Equation \ref{eq:factor},
we generically require $\Omega(d^{2n-4})=\Omega(d^{2n})$ gates of the
two-qudit library to realize symmetry-less unitary evolutions within
$U(d^{n})$.

A similar invocation of Sard's theorem
produces a lower bound on circuit sizes for state
synthesis.  Here, the problem is to produce the most efficient
possible circuit $U$ capable of realizing
generic $\ket{\psi} \in \mathcal{H}(n,d)$ from a fixed start-state
typically chosen to be $\ket{0}$, i.e. building a small
circuit for $U$ so that $\ket{\psi}=U \ket{0}$.
We claim that circuits for generic state-synthesis require
$\Omega(d^n)$ gates.  Indeed, $U\ket{0}$ is simply the first column
of the matrix realization of $U$, and taking the column of a matrix
is a smooth map.  Hence we apply the Sard's theorem argument above
to $\tilde{f}:U(d^2)^\ell \rightarrow \mathcal{H}(d,n)$,
whence the result.  Now a subcircuit of our universal unitary
qudit-evolution circuit, described in 
Equation \ref{eq:household_state}, is also capable of solving the state
synthesis problem in $(d^n-1)/(d-1) \in O(d^n)$ two-qudit gates. 
Hence the qudit state-synthesis generically requires $\Theta(d^n)$ gates.

\bigskip

\noindent
{\bf Theorem:} \ \ 
\emph{  The following asymptotics hold for $d$ multi-level quantum logic
circuits.  In each statement, $d$ is fixed and the asymptotic is
stated exclusively in terms of $n$.
\begin{enumerate}
\item
Given a generic $\ket{\psi}$, constructing a quantum circuit
for a unitary $U$ such that $U \ket{0}=\ket{\psi}$ requires $\Theta(d^n)$
two-qudit gates.
\item
Constructing a quantum circuit for a generic $n$-qudit unitary operator
$U \in U(d^{2n})$ consisting of two-qudit gates requires
$\Theta(d^{2n})$ two-qudit gates.
\end{enumerate}
}

\bigskip

As a remark, other gate libraries that are asymptotically equivalent
to two-qudit gates might be better suited to certain problems.
In reasonable cases, there should be a fixed upper bound on the number
of library-gates required to realize a two-qudit unitary operator.
For any such library a sharp asymptotic of $\Theta(d^{2n})$ gates is likewise 
required for generic unitary evolution.  This holds in particular
for \hbox{$\{\mbox{local unitary}\} \sqcup
\{ \bigwedge_1( \sigma^x \oplus I_{d-2})\}$} \cite{BrennenOLearyBullock:04}.

\section{Qubit Emulation is Insufficient}
\label{sec:emulate}

Consider two emulation schemes of qudits by qubits:
\begin{enumerate}
\item 
\label{it:individual}
One might emulate each individual qudit
with as few qubits as possible, so that the local qudit structure is
respected.
\item 
\label{it:pack}
One might rather pack the entire $d^n$ dimensional $n$-qudit
state into the smallest possible qubit state space, ignoring the
local (tensor) structure.  
\end{enumerate}
We argue that the emulation circuit for
Option \ref{it:individual} 
does not attain the lower bound asymptotic, while in essense
Option \ref{it:pack} does not allow for circuit-level
emulation at all.  

In Option \ref{it:individual},
label $\beta=\lceil \log_2 d \rceil$, so that $\beta$ qubits are required to
emulate a qudit.  Now for the qubit circuit diagram, some multi-qubit
gates will in fact be local to the qudit, while others are genuine
two-qudit gates.  Hence, if $U$ is a $d^{n} \times d^{n}$ 
unitary matrix and the $O(2^{2 \beta n})$ circuit is applied after splitting
each qudit into $\beta$ virtual qubits, we obtain an \emph{upper}
bound of $O(2^{2 \beta n})$ two-qudit gates.  Note that this asymptotic
is worse than both $O(d^{2n})$ and even $O(n^k d^{2n})$ unless
$d$ is a power of two.
(For $2^{2 \beta} > d^2$ in this case, so the exponentials have distinct
bases and are not asymptotically equivalent.)
Thus, prior art does not suffice
for the upper bound asymptotic.

We next consider Option \ref{it:pack}.  Note that
$n$ qudits may be viewed as $d^n$ Hilbert space dimensions.
Ignoring the
local structure, a unitary evolution of $\mathcal{H}(n,d)$ may be realized
as a subblock of a unitary evolution of $n \delta = 
\lceil n \log_2 d \rceil$ qubits
rather than $n\beta=n \lceil \log_2 d \rceil$ as above.  Indeed, 
with this form of emulation,
it is true that $O(4^{n\delta})=O(d^{2n})$ 
virtual two-qubit gates would suffice by earlier methods.
However, in this mode of emulation a virtual two-qubit gate
\emph{need not} correspond to a two-qudit gate.  Indeed, it might not
even be a $k$-qudit gate for $k$ small.  Consider for example two-qutrit
gate of the form $I_3 \otimes V$ acting on $\mathcal{H}(3,3)$, where
$V \in U(3^2)$.  
This has a $9 \times 9$ block structure,
but emulating such a unitary using qubits is more or less an arbitrarily
difficult 
$5$-qubit evolution.  It is certainly not a two-qubit gate!
Thus, although the mapping between Hilbert spaces is possible,
tensor (Kronecker) product structures are not preserved.

There are several candidate systems for quantum computation where the 
physical subsystems encoding the quantum information 
have dimension $d>2$. Examples include charge-position states in 
quantum dots \cite{SchirmerEtAl:03}, 
rotational and vibrational states of a molecule \cite{ShapiroEtAl:03}, 
truncated subspaces of harmonic oscillator states \cite{BartlettEtAl:02} 
and ground electronic states of alkali atoms with total 
spin $F>1/2$ \cite{KloseEtAl:01}. Moreover,
it is useful to allow $d$ not a power of two. First, in many instances, 
the physics of the system can preclude encoding in a Hilbert space of 
arbitrary size. For example, in the case of encoding in 
alkali atoms the Hilbert space dimension of a single hyperfine 
manifold is $2F+1$ so for bosonic atoms, $d$ is never a power of two.
\footnote{The dimension of the total ground state Hilbert space including 
both manifolds corresponding to the two spin states of the 
valence electron may be a power of two e.g. $^{87}$Rb 
and $^{133}$Cs.} Second, there is evidence that the fault-tolerant 
threshold for quantum computation can be improved when using error 
correction codes on qudits with $d$ prime \cite{Aharonov:01}.

\section{Controlled One-Qudit Operators}
\label{sec:uniform_control}

Although the complexity bound is phrased in terms of two-qudit operators,
our $QR$ factorization algorithm in \S \ref{sec:upper}
produces a quantum circuit of operators that act on one target 
qudit depending on the state of multiple control qudits.  The majority of the
$k$-controlled one-qudit operators are doubly $(k=2)$
or singly $(k=1)$ controlled.  We next review 
how a $k$-controlled qudit operator may be realized in
$O(k)$ two-qudit gates, given $r=\lceil (k-1)/(d-2) \rceil$ ancilla qudits.

\subsection*{Qudit Generalizations of {\tt CNOT}}

The most common two-qubit
gate is the quantum controlled-not, due to its appearance in early
papers on quantum computing and reversible
classical computation.  In case $d=2$, this gate, denoted 
{\tt CNOT} or $\bigwedge_1(\sigma^x)$, linearly
extends the action of the classical {\tt CNOT} on bit-strings to
two-qubit kets.  Thus {\tt CNOT} applies a {\tt NOT} ($\sigma^x$)
iff the control bit is in state $\ket{1}$.
So in two qubits with control on the most significant qubit, {\tt CNOT}
linearly extends $\ket{00}\mapsto \ket{00}$, $\ket{01} \mapsto \ket{01}$,
$\ket{10} \mapsto \ket{11}$, and $\ket{11}\mapsto \ket{10}$.  An extension
to arbitrary $d$ has been suggested \cite{BrennenOLearyBullock:04}. 
We may view $\sigma^x$ as addition mod $2$, 
which generalizes as follows.  
If $c \in \mathbb{Z}/d\mathbb{Z}$ is a dit, then
we use $(\oplus c)$ to (abusively) denote both the addition map
$k \mapsto k \oplus c$ within $\mathbb{Z}/d\mathbb{Z}$ and also the
one-qudit unitary operator given by the permutation matrix of this map.
So for example in qutrits ($d=3$,)
\begin{equation}
\oplus 1 \ = \ {\tt INC} \ = \ \left( \begin{array}{rrr} 0 & 0 & 1 \\
1 & 0 & 0 \\ 0 & 1 & 0 \\ \end{array} \right)
\end{equation}
A corresponding
(unitary) permutation map {\tt INC} is given by
${\tt INC} \ket{j} = \ket{(j+1) \mbox{mod } d}$ for any base $d$.

Then the {\tt CINC} (controlled-increment) gate applies 
{\tt INC} iff the control qudit is in state $\ket{d-1}$, i.e.
in the case of most-significant qudit control
\begin{equation}
{\tt CINC} \ket{j,k} \ = \ 
\left\{
\begin{array}{rr}
\ket{j, (k+1) \mbox{mod } d} & j = d-1 \\
\ket{j,k} & j\neq d-1 \\
\end{array}
\right.
\end{equation}
We take the $\oplus$
symbol in a circuit diagram to designate modular increment {\tt INC} as in the
$d=2$ case, so that the usual symbol for {\tt CNOT} in
a $d$-level diagram now designates {\tt CINC}.

Using the most recent argument \cite{BrennenOLearyBullock:04}, a second
generalization of {\tt CNOT} must be added to the qudit local unitary
group $U(d)^{\otimes n}$ in order to recover exact universal qudit
computation.  For this, we label $\sigma^x \oplus I_{d-2}$ as that
computation with $\ket{0} \leftrightarrow \ket{1}$ and
all other $\ket{j} \mapsto \ket{j}$, $2 \leq j \leq d-1$.  Then
the appropriate second gate is $\bigwedge_1(\sigma^x \oplus I_{d-2})$.  
Note that a {\tt CINC} gate may be constructing
using {\tt INC} gates and $d-1$ copies of
$\bigwedge_1(\sigma^x \oplus I_{d-2})$ \cite{BrennenOLearyBullock:04}.
Since a $QR$ argument ibid. also produces any two-qudit operator
with at most $O(d^2)=O(1)$ gates from the library
\hbox{$\{\mbox{local unitary}\} \sqcup 
\{ \bigwedge_1( \sigma^x \oplus I_{d-2})\}$}, the optimal asymptotics
of the Theorem apply equally well to this library.

\subsection*{Emulation of multiple-controlled operations by single control operations}

The complexity of a quantum algorithm is determined
by the asymptotic number of single qudit and two-qudit gates necessary to implement the 
corresponding unitary.
We describe here how to emulate controlled one-qudit operations using
only single qudit and two-qudit gates.  In $n$ qudits, a controlled one-qudit operator $V$ is applied to a 
target qudit based on a string of $n-1$ controls.
Each control is either $*$, to denote a match with
an arbitrary value (no control,) 
or is chosen to be one of $0,1,\dots,d-1$, 
to force a specific matching value (control.)
Note that single qudit control and local operations may be used to
emulate multiple-controlled gates at low cost.  In circuit diagrams, we will
denote a control triggering on 
an arbitrary state $\ket{j}$ with a box, in contrast to the standard
control denoted by a bullet that only triggers on state $\ket{d-1}$.
One formal definition of a controlled one-qudit gate is the following:

\begin{definition}[Controlled one-qudit operator 
     $\bigwedge({C},V)$]
\label{def:uniform_control}
Let $V$ be a $d \times d$ unitary matrix, i.e. a one-qudit operator. 
Let ${C}=[ {C}_1 {C}_2 \ldots {C}_n]$ 
be a length-$n$ \emph{control word} 
composed of letters from the alphabet
$\{0,1,\ldots,d-1\} \sqcup \{ \ast \} \sqcup \{T\}$,
with exactly one letter in the word being $T$.
By $\#{C}$ we mean the number of letters in the word
with numeric values (i.e., the number of controls,) 
and the set of control qudits is the corresponding
subset of $\{1,2,\ldots,n\}$ denoting the positions of numeric
values in the word.
We will say that a control word \emph{matches} an $n$-dit string
if each numeric value matches.
Then the controlled one-qudit operator 
$\bigwedge({C},V)$
is the $n$-qudit operator that applies $V$ to the qudit specified
by the position of $T$
iff the control word matches the data state's $n$-dit string.
More precisely,
in the case when ${C}_n=T$, then
\begin{equation}
\bigwedge( [{C}_1 {C}_2 \ldots {C}_{n-1} T],V) 
\ket{c_1 c_2\ldots c_n} \ = \ 
\left\{
\begin{array}{rr}
\ket{c_1 \ldots c_{n-1}} \otimes V\ket{c_n}, & c_j = {C}_j 
\mbox{ or } {C}_j=\ast, \ 1 \leq k \leq n-1 \\
\ket{c_1 \ldots c_{n-1} c_n}, & \mbox{else} \\
\end{array}
\right.
\end{equation}
Alternatively, if ${C}_j=T$ ($j < n$,) 
we consider the unitary (permutation)
operator $\chi_j^n$ that swaps qudits $j$ and $n$.  Thus, 
$\chi_j^n \ket{d_1 d_2 \ldots d_n} =
\ket{d_1 d_2 \ldots d_{j-1} d_n d_{j+1} \ldots d_{n-1} d_j}$.
We remark that $\chi_j^n = (\chi_j^n)^T = (\chi_j^n)^{-1}$.
We apply the same permutation to
${C}= [ C_1 C_2 \ldots C_{j-1} T C_{j+1} \ldots C_n ]$,
obtaining
$\tilde{{C}}=[C_1 C_2 \ldots C_{j-1} C_n C_{j+1} \ldots C_{n-1} T]$
and we 
define $\bigwedge({C},V)= \chi_j^n \bigwedge(\tilde{{C}},V)
\chi_j^n$.
\end{definition}

\begin{figure}[t]
\begin{center}
\[
\Qcircuit @R 0.5em @C 0.5em {
& \controlu \qw & \qw &  & & \qw &  \gate{\oplus^{d-d_1-1}} & \ctrl{1} 
& \gate{\oplus^{d_1+1}} & \qw & & &  \gate{\oplus^{d-d_1-1}} & \ctrl{5} 
& \qw & \qw & \qw & \qw & \qw & \qw & \qw & \qw & \qw & \ctrl{5} 
& \gate{\oplus^{d_1+1}} & \qw & \\
& \controlu \qw \qwx & \qw & & & \qw & \gate{\oplus^{d-d_2-1}} & \ctrl{1}
& \gate{\oplus^{d_2+1}} & \qw & & &  \gate{\oplus^{d-d_2-1}} & \qw &
\ctrl{4} & \qw & \qw & \qw & \qw & \qw & \qw & \qw & \ctrl{4} & \qw 
& \gate{\oplus^{d_2+1}} & \qw & \\
& \controlu \qw \qwx & \qw & \ustick{{\buildrel \cong \over {\ }}} &  & 
\qw & \gate{\oplus^{d-d_3-1}} & \ctrl{1} & \gate{\oplus^{d_3+1}} & \qw 
& \ustick{{\buildrel \cong \over {\ }}} &  &  \gate{\oplus^{d-d_3-1}} & \qw 
& \qw & \qw & \ctrl{4} & \qw & \qw & \qw & \ctrl{4} & \qw & \qw & \qw 
& \gate{\oplus^{d_3+1}} & \qw & \\
& \gate{V} \qwx & \qw  & & & \qw & \qw & \gate{V} & \qw  & \qw 
& & & \qw & \qw & \qw & \qw & \qw & \qw & \gate{V} & \qw & \qw & \qw & \qw 
& \qw & \qw & \qw & \\
\\
& & & & & & & &  & & & & \lstick{\ket{0}} & \targ \qw & \targ & \ctrl{1} & \qw
& \qw & \qw & \qw & \qw & \ctrl{1} & \targ & \targ & \qw & \qw & 
\rstick{\ket{0}}\\
& & & & & & & &  & & & & \lstick{\ket{0}} & \qw & \qw  & \targ & \targ &
\qw & \ctrl{-3} & \qw & \targ & \targ & \qw & \qw & \qw & \qw 
& \rstick{\ket{0}} \\
}
\]
\end{center}
\caption{
\label{fig:wedgek}
Qutrit ($d=3$) emulation 
of $\bigwedge([d_1 d_2 d_3 T], V)$ using at most two-qudit operations.
The general argument shows that $\bigwedge({C},V)$ requires at
most $O(\# {C})$ two-qudit gates.  The rightmost circuit diagram
is well-known \cite[Figure 1]{MuthukrishnanStroud:00}.
Also, the controlled-V
may be decomposed into local operators, {\tt CINC}s,
and $\bigwedge_1(\sigma^x \oplus I_{d-2})$ if the finer gate library
is of interest \cite{BrennenOLearyBullock:04}.
}
\end{figure}
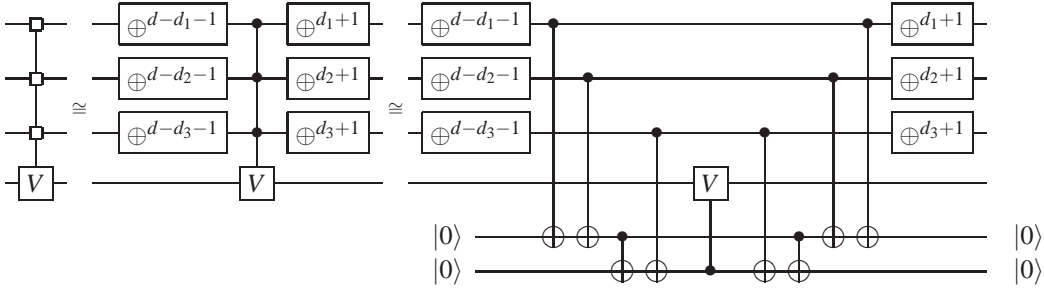

We note an earlier 
simulation \cite{MuthukrishnanStroud:00}
of $\bigwedge({C},V)$ in terms of two-qudit operations.
In addition to $n$ data qubits, we also require 
$r=\lceil (n-2)/(d-2) \rceil$ \cite{MuthukrishnanStroud:00}
ancilla qudits initially set to $\ket{0}$, as illustrated 
in Figure \ref{fig:wedgek} for
$\# {C}=3$ and qutrits ($d=3$.)
The idea is to use local operations to control on any logical 
basis state.  Then a sequence of {\tt CINC}'s appropriately
targeting the $r$ ancillas change the state of the last ancilla
to $\ket{d-1}$ if and only if each control line carries $\ket{d-1}$.
The entire operation follows by applying a singly-controlled $V$
using this last ancilla and then mirroring the {\tt CINC} pattern
in order to disentangle the ancilla qudits from the data qudits.

\section{Asymptotically Optimal Qudit State-Synthesis}
\label{sec:Householder}

The key component of our universal qudit circuit
is a subcircuit interesting in its own right:  given a state vector
$\ket{\psi} \in \mathcal{H}(n,d) \cong \mathbb{C}^{d^n}$, we construct a
sequence of $p$ controlled one-qubit operators depending on
$\ket{\psi}$ such that
\begin{equation}
\label{eq:household_state}
\prod_{k=1}^{p} \bigwedge[{C}(p-k+1),V(p-k+1)] \; \ket{\psi}
\ = \ \ket{0} \, .
\end{equation}
We remark that we use $V(p-k+1)$ instead of $V_{p-k+1}$, since the
latter sort of subscript is often used to denote a target qudit while
we intend the target to be labelled by the $T$ symbol within 
the word ${C}(p-k+1)$.

Before continuing to
construct this component of our universal circuit, we note that this
subcircuit achieves asymptotically optimal qudit state-synthesis.
The state-synthesis problem is to construct efficient (small) quantum
circuits whose associated unitary $U$ has $U \ket{0} = \ket{\psi}$ for
some arbitrary but pre-determined $\ket{\psi} \in \mathcal{H}(n,d)$.
For $d=2$, several works address this topic, e.g.
\cite{Mottonenb:04,ShendeBullockMarkov:04,Knill_state,Deutsch_state}.
For our subcircuit in Equation \ref{eq:household_state},
note that
\begin{equation}
\label{eq:state_synth}
\prod_{k=1}^{p} \bigwedge [{C}(k),V(k)^\dagger]\; \ket{0} \ = \ 
\ket{\psi}
\end{equation}
Our construction realizes any $\ket{\psi}$ in 
$p=(d^n-1)/(d-1) \in O(d^n)$ two-qudit gates,
since also $\# \mathcal{C}(k) \leq 1$ for all $k$.  Given the $\Omega(d^n)$
lower bound of the introduction, 
we conclude that qudit state
synthesis generically requires $\Theta(d^n)$ gates.

Finally, we briefly note our decision to index the sequence
of $\bigwedge({C},V)$ so that the earlier indices appear
on the right.  There are two reasons for this.  First, it means
the index describes the operators in the order in which they
are applied to $\ket{\psi}$, rather than the reverse.  Second,
the state-synthesis has received more attention than generalized
Householder reductions in the literature, and note that the indices
of Equation \ref{eq:state_synth} do increase to the right.

\subsection*{One-qudit Householder Reflections}

Earlier universal $d=2$ circuits
\cite{BarencoEtAl:95} relied on a $QR$ factorization to write
any unitary $U$ as a product of \emph{Givens rotations}, realized
in the circuit as $k$-controlled unitaries \cite{Cybenko:01}.  
Such Givens rotations $V$
coincide with the identity matrix except in the pairwise intersection
of rows $j$, $k$,
with columns $j$, $k$.  Here, the entries 
$V_{jj}, V_{jk}, V_{kj}, V_{kk}$ entries mimic those of a
$2\times 2$ unitary matrix.  
Thus, a Givens rotation is geometrically a rotation in the $jk$-plane.  
In the multi-level case, we use 
\emph{Householder reflections} \cite[\S5.1]{gvl}
instead of Givens rotations, in order to take full advantage of
the range of single qudit operators.

Thus, suppose $\ket{\psi} \in \mathcal{H}(1,d)$, perhaps not
normalized, and suppose we wish to construct a unitary operator
$W$ such that $W \ket{\psi}$ is a multiple of $\ket{0}$.  Standard
formulas exist for constructing such $W$ for real vectors.  For a 
complex vector, these formulas become
\begin{equation}
\label{eq:eta_equation}
\left\{
\begin{array}{lcl}
\ket{\eta} & = & \ket{\psi}-\sqrt{\langle \psi | \psi \rangle}
\frac{ \langle    0 | \psi \rangle} {\big| \langle     0| \psi \rangle \big| }
\ket{0} \\
W & = & I_d - (2/\langle \eta | \eta \rangle) \; \ket{\eta}\bra{\eta} \\
\end{array}
\right.
\end{equation}
Then indeed $W \ket{{\psi}}$ is a multiple of $\ket{0}$.

\subsection*{$n$-qudit State-Synthesis}

We next describe the algorithm for realizing
$\prod_{k=1}^{p} \bigwedge[{C}(p-k+1),V({p-k+1})] \; \ket{\psi}
\ = \ \ket{0}$ with $\# {C}(k) \leq 1$ and $p=(d^n-1)/(d-1)$.
The circuit topology has a recursive structure that we abstract
into the following algorithm that generates
the $\clubsuit$-sequence (``club-sequence".)

\begin{figure}[t]
\begin{tabular}{||c|l||}
\hline 
$n$ & $\clubsuit$-sequence, $d=3$ \\
\hline
\hline
$1$ & $\clubsuit$ \\
\hline
$2$ & $0 \clubsuit$, $1 \clubsuit$, $2 \clubsuit$, $\clubsuit \clubsuit$ \\
\hline
$3$ & $0 0 \clubsuit$, $01\clubsuit$, $02 \clubsuit$, $0 \clubsuit \clubsuit$,
$10 \clubsuit$, $11 \clubsuit$, $12 \clubsuit$, $1 \clubsuit \clubsuit$,
$2 0 \clubsuit$, $2 1 \clubsuit$, $2 2 \clubsuit$, $2 \clubsuit \clubsuit$,
$\clubsuit \clubsuit \clubsuit$ \\
\hline
$4$ & $000 \clubsuit$, $001\clubsuit$, $002 \clubsuit$, 
$00 \clubsuit \clubsuit$,
$010 \clubsuit$, $011 \clubsuit$, $012 \clubsuit$, $01 \clubsuit \clubsuit$,
$020 \clubsuit$, $021 \clubsuit$, $022 \clubsuit$, $02 \clubsuit \clubsuit$,
$0\clubsuit \clubsuit \clubsuit$ \\
& $100 \clubsuit$, $101\clubsuit$, $102 \clubsuit$, 
$10 \clubsuit \clubsuit$,
$110 \clubsuit$, $111 \clubsuit$, $112 \clubsuit$, $11 \clubsuit \clubsuit$,
$120 \clubsuit$, $121 \clubsuit$, $122 \clubsuit$, $12 \clubsuit \clubsuit$,
$1\clubsuit \clubsuit \clubsuit$ \\
& $200 \clubsuit$, $201\clubsuit$, $202 \clubsuit$, 
$20 \clubsuit \clubsuit$,
$210 \clubsuit$, $211 \clubsuit$, $212 \clubsuit$, $21 \clubsuit \clubsuit$,
$220 \clubsuit$, $221 \clubsuit$, $222 \clubsuit$, $22 \clubsuit \clubsuit$,
$2\clubsuit \clubsuit \clubsuit$, $\clubsuit \clubsuit \clubsuit \clubsuit$ \\
\hline

\end{tabular}
\caption{  
\label{fig:clubsuit}
Sample $\clubsuit$-sequences for $d=3$, i.e. qutrits.}
\end{figure}

\bigskip
 
\vbox{

\noindent
{\bf \hrule}
\smallskip

\noindent
Algorithm 1:  $\{ s_{1},\dots,s_p\}$ 
= {\bf Make-$\clubsuit$-sequence}$(d,n)$

\smallskip

\noindent
{\hrule}

\smallskip

\begin{tabular}{l}
\%  {\em We return a sequence of $p=(d^n-1)/(d-1)$ terms, with
$n$ letters each,} \\
\% {\em drawn from the alphabet $\{ 0,1,\dots,d-1,\clubsuit \}$.} \\
Let $\{ \tilde{s}_{j}\}_{j=1}^{\tilde{p}}$ =  
         {\bf Make-$\clubsuit$-sequence} ($d$,$n-1$.) \\
{\bf for } $q=0,1,\dots,d-1$ {\bf do} \\
\quad The next $(d^{n-1}-1)/(d-1)$ terms of the sequence
are formed by prefixing the letter $q$ to each \\
\quad term
of the sequence $\{ \tilde{s}_{j}\}$.\\
{\bf end for} \\
The final term of the sequence is $\clubsuit^n$. \\
\end{tabular}

\noindent
{\hrule}
}

\bigskip

Sample $\clubsuit$-sequences that illustrate the construction 
are given in Figure \ref{fig:clubsuit}.  Note that the number of
elements in the sequence equals the number of
uncontrolled or singly-controlled one-qudit operators in the
state-synthesis circuit.  
We choose to describe the circuit by iterating over the sequence.
Thus, in order to produce the circuit,
it suffices to describe how to extract the control word ${C}$ from a
term $t$ of the $\clubsuit$-sequence and how to determine $V$ from
the term and $\ket{\psi_j}$, where 
$\ket{\psi_j}=
\prod_{k=1}^{j-1} \bigwedge[{C}(p-k+1),V({p-k+1})] \; \ket{\psi}$
is the partial product.  This may be done as follows.

\bigskip
 
\vbox{
\noindent
{\bf \hrule}

\smallskip

\noindent
Algorithm 2:  \ \ $\bigwedge({C},V)$ = {\bf Single-$\clubsuit$Householder}
($\clubsuit$ term $t=t_1t_2 \ldots t_n$, \,  $n$-qudit state $\ket{\psi_j})$

\noindent
{\hrule}

\smallskip

\begin{tabular}{l}
Initialize ${C} = \ast \ast \cdots \ast$ \\
\% {\em Set the target:} \\
Let $\ell$ be the index of the leftmost $\clubsuit$
      and set $C_{\ell} = T$. \\
\% {\em Set a single control if needed:} \\
{\bf if} $t$ contains numeric values greater than 0, \\
\quad Let $q$ be the index of the rightmost such value
      and set $C_q = t_q$. \\
{\bf end if} \\
Given $\ket{\psi_j}=\sum_{k=0}^{d^n-1} \bra{k} \psi_j \rangle \ket{k}$,
form a one-qudit state $\ket{\varphi}=\sum_{k=0}^{d-1} \bra{t_1 t_2 \ldots
t_{\ell-1} k 0 0 \ldots 0} \psi_j \rangle \ket{k}$. \\
Form $V$ as one-qudit Householder such that  $V\ket{\varphi}=\ket{0}$. \\
\end{tabular}

\noindent
{\hrule}
}

\bigskip

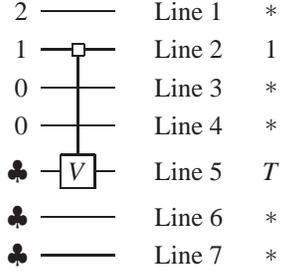
\begin{figure}[t]
\begin{center}
\[
\Qcircuit @R 0.75em @C 0.75em {
& \lstick{2}  & \qw & \qw & & \push{\mbox{Line }1} && \push{\ast} \\
& \lstick{1}  &  \controlu \qw & \qw &&  \push{\mbox{Line }2} 
&& \push{1} \\
& \lstick{0}  & \qw \qwx & \qw &&  \push{\mbox{Line }3} && \push{\ast} \\
& \lstick{0} & \qw \qwx & \qw &&  \push{\mbox{Line }4} && \push{\ast} \\
& \lstick{\clubsuit} & \gate{V} \qwx & \qw && \push{\mbox{Line }5} 
&& \push{T} \\
& \lstick{\clubsuit} & \qw & \qw &&  \push{\mbox{Line }6} && \push{\ast} \\
& \lstick{\clubsuit} &  \qw & \qw && \push{\mbox{Line }7} && \push{\ast}
}
\]
\end{center}
\caption{
\label{fig:club_to_control}
Producing a $\bigwedge({C},V)$ given $V$ and a term of the
$\clubsuit$-sequence, here $t = 2100\clubsuit \clubsuit \clubsuit$ for
seven qudits.  The algorithm for producing ${C}$ places
the $V$-target symbol $T$ on the leftmost club, here line
$5$.  The active control must then be placed on the least
significant line carrying a nonzero prior to line $5$, here the $1$
on line 2.  (A control on lines 3 or 4 would not
prevent the nonzero $\alpha_0$ of $\ket{\psi_j} = \sum_{k=0}^{d^n-1}
\alpha_k \ket{k}$ from creating new nonzero entries in previously
zeroed positions.)  Thus in this case, 
${C} = \ast 1 \ast \ast T \ast \ast$.  The $V$ is
chosen to zero all but one $\alpha_k$ for $k=2100\ell 00$.
}
\end{figure}

\begin{figure}[t]
\begin{center}
\includegraphics[scale=0.25]{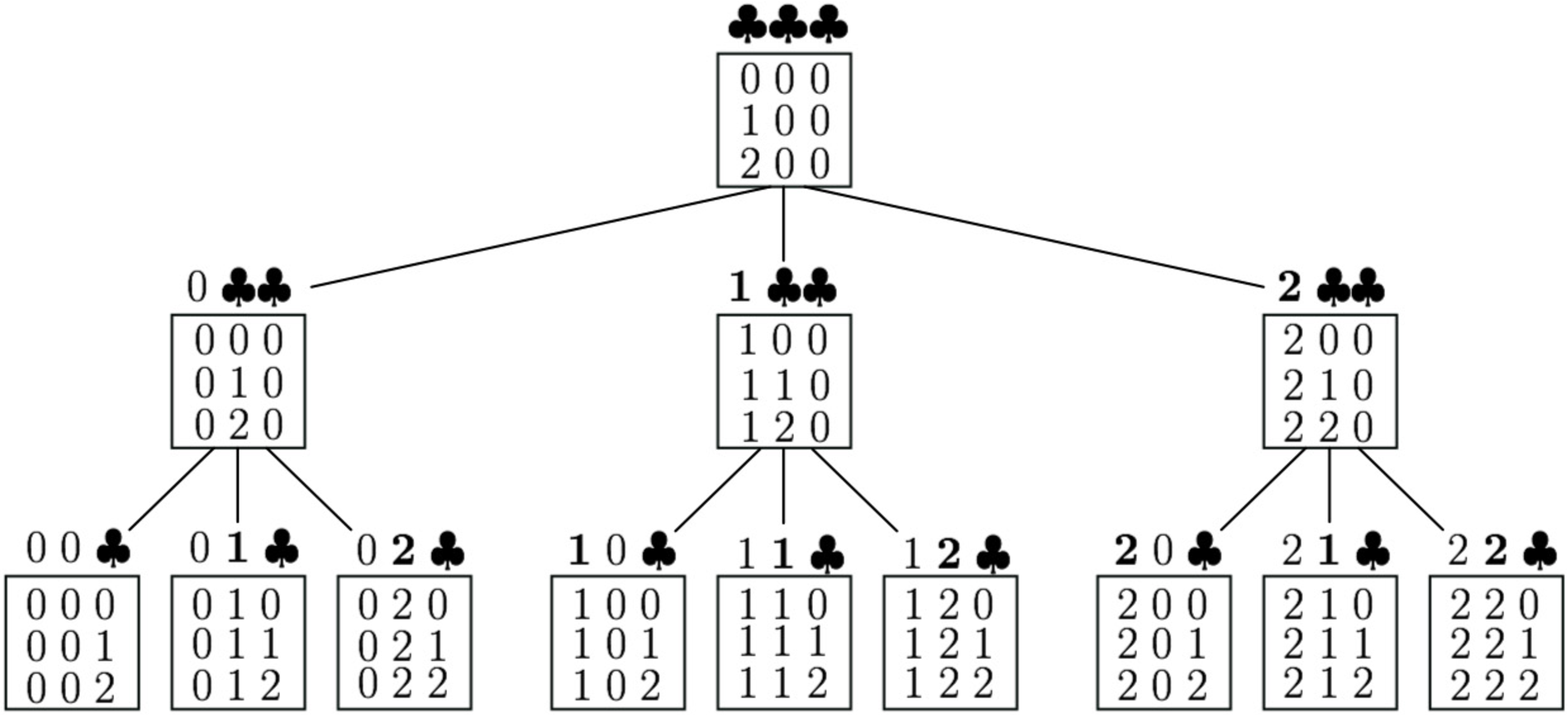}
\end{center}
\caption{
\label{fig:Housegraph}
Using the $\clubsuit$-sequence for $d=3$, $n=3$ to generate
Householder refections to reduce $\ket{\psi}$ to a multiple of $\ket{0}$.
Each node is labeled by a $\clubsuit$-term
and represents a Householder reflection $\bigwedge(C,V)$.
The control is indicated by the boldface entry in the label.
As the tree is traversed in a depth-first search, each node indicates a
$\bigwedge(C,V)$ which zeroes the components of the
last two indices in each node using the component of the top entry.
}
\end{figure}

Figure \ref{fig:club_to_control} illustrates the gate
produced from the output $C$ and $V$ from 
the algorithm {\bf Single-$\clubsuit$Householder}.
Figure \ref{fig:Housegraph} illustrates the order in which these
$\bigwedge(C,V)$
reflections are generated if we iterate over the
$\clubsuit$-sequence.  Each node of the tree
is labeled by a $\clubsuit$-term
and represents a Householder reflection
defined by the three indicated elements of $\ket{\psi}$.  
After the reflection,
the first element in the node
remains and the others are zeroed.  The reflections
are applied by traversing the graph in depth-first order, left to
right. 
To understand the controls, notice that the leftmost Householder
on each layer of the graph requires no control.  For example,
the Householder $0 \clubsuit \clubsuit$ defined by elements
$0j0$ ($j = 0,1,2$) is applied to 9 sets of elements:
$0j1$ and $0j2$, all zeroed, and
$1j0, 1j1, 1j2, 2j0, 2j1, 2j2$, as yet not zeroed.
For the other Householder nodes, the control is indicated in
boldface.  
For the leftmost Householder in a group of $d$ siblings,
we do not wish to touch elements in groups to the left of it.  
So we set the control to stay within the group.  For
example, the Householder labeled $10\clubsuit$ is also applied to 
elements in $11\clubsuit$ and $12\clubsuit$.
For other Householders in a group, the corresponding elements in
groups to the left are completely zero, and in groups to the
right are as yet unzeroed.  Thus, for example, the Householder
labeled $11\clubsuit$ is applied to $01\clubsuit$ (all zero
since $0\clubsuit\clubsuit$ has already been applied) and
$21\clubsuit$.
We can formalize this argument to a proof of correctness as given in
Appendix \ref{app:correct}.

\subsection*{Householder Circuits Retaining $\ket{j} \neq
\ket{0}$}

In the $QR$ unitary-circuit application, we will need not only
$\ket{\psi} \mapsto \sqrt{\bra{\psi} \psi \rangle} \ket{0}$
but also 
$\ket{\psi} \mapsto \sqrt{\bra{\psi} \psi \rangle} \ket{j}$
for any $j=d_1 d_2 \ldots d_n$.  Rather than provide a new
algorithm, we instead adapt our
algorithm for a collapse onto $\ket{0}$ into an algorithm
for collapse onto $\ket{j}$.  
The idea is to permute the elements to put $j$ in position $0$, 
apply {\bf Single-$\clubsuit$Householder},
and then permute back.
The rest of this subsection describes this in detail.

We abusively continue to use $\oplus p$
for $1 \leq p \leq d-1$ to denote the one-qudit unitary operator that
carries $\ket{k} \mapsto \ket{(p+k) \mbox{mod } d}$, i.e.
$\oplus p = {\tt INC}^p$.  Given the $d$-ary expansion of $j$,
we have $\otimes_{k=1}^n [\oplus d_k] \ket{00\ldots 0} = \ket{j}$.
Consider $\bigwedge({C},V)$, and define  
$\tilde{{C}}$ by
\begin{equation}
\tilde{{C}}_k =
\left\{
\begin{array}{rr}
\ast, & {C}_k = \ast \\
T, & {C}_k = T \\
({C}_k+d_k) \mbox{mod } d, & 
{C}_k \in \{ 0,1,\dots,d-1\}      \\
\end{array}
\right.
\end{equation}
Suppose also that ${C}_\ell = T$.
Then noting that $(\oplus j)^\dagger = \oplus (d-j)$, we have
the similarity relation
\begin{equation}
\label{eq:flip_plus}
[\otimes_{k=1}^n \oplus(d_k)] \bigwedge({C},V) [\otimes_{k=1}^n
\oplus (d-d_k)] \ = \ 
\bigwedge[\tilde{{C}}, (\oplus d_\ell) V (\oplus d-d_\ell)]
\end{equation}
This is the basis for the algorithm.

\bigskip

\vbox{
\noindent
{\hrule}
\noindent
Algorithm 3:  \ \ $\bigwedge({C},V)$ = {\bf $\clubsuit$Householder}
$(\ket{\psi},j,d,n)$

\noindent
{\hrule}

\smallskip

\begin{tabular}{l}
\% {\em Reduce $\ket{\psi}$ onto $\ket{j}$.} \\

Let $j=d_1d_2\ldots d_n$. \\
Compute $\ket{\varphi}
\ =\ [\otimes_{q=1}^n \oplus d-d_q]\; \ket{\psi}$. \\
Produce a sequence of controlled one-qudit operators so that \\
\quad $\prod_{k=1}^p \bigwedge[{C}(p-k+1),V(p-k+1)] 
      \ket{\varphi}=\ket{00\ldots 0}$,\\
\quad  using {\bf Single-$\clubsuit$Householder}
      applied to each term of {\bf Make-$\clubsuit$-sequence}$(d,n)$. \\
Compute
      $[\otimes_{q=1}^n ( \oplus d_q) ]\bigwedge[{C}(p-k+1),V(p-k+1)]
      [\otimes_{p=1}^n (\oplus \; d-d_q)]=$ \\
\quad \quad $\bigwedge[\tilde{C}(p-k+1), \tilde{V}(p-k+1)]$ 
      using Equation \ref{eq:flip_plus}\\
\end{tabular}

\noindent
{\hrule}
}

\section{A Qudit-native $QR$-based Quantum Circuit Synthesis Algorithm}
\label{sec:upper}

The asymptotically optimal qudit-universal circuit we present does
not require Gray codes (Cf. \cite{Vartiainen:04}.)  Rather, it
leans heavily on the optimal state-synthesis of \S \ref{sec:Householder}.
Since this state-synthesis circuit can likewise clear any
length $d^n$ vector using fewer than $d^n$ single controls, the
asymptotic is perhaps unsurprising.  However, the recursive nature
of our synthesis algorithm requires highly-controlled one-qudit unitary
operators when clearing entries near the diagonal, and other highly-controlled
one-qudit unitaries are needed to finish clearing each column.  In
presenting the algorithm, we highlight two themes:
\begin{itemize}
\item  We process the size $d^n \times d^n$ unitary $V$ in subblocks of
size $d^{n-1} \times d^{n-1}$.
\item  Due to rank considerations, at least one block in each
block-column of size $d^{n}\times d^{n-1}$ must remain full rank throughout.
\end{itemize}
Hence, we cannot carelessly zero subcolumns.  One solution is to
triangularize the $d^{n-1} \times d^{n-1}$ matrices
on the block diagonal, recursively.  We also note that only $O(n^2d^n)$
fully ($n-1$) controlled one-qudit operations appear in the algorithm, 
which is allowed when working towards an asymptotic of 
$O(d^{2n})$ controls total.

The organization for the algorithm is then as follows.  Processing
(triangularization) of $V$ moves along block-columns of
size $d^n \times d^{n-1}$ from left to right.  In each block-column,
we first triangularize the block $d^{n-1} \times d^{n-1}$ block-diagonal
element, perhaps adding a control on the most significant qudit
to a circuit produced by recursive triangularization.  After this recursion,
we zero the blocks below the block-diagonal element one column
at a time.  For each column $j$, $0 \leq j \leq d^{n-1}-1$, the zeroing
process is to collapse the $d^{n-1} \times 1$ subcolumns onto their
$j^{\mbox{th}}$ entries, again adding a 
control on the most significant
qudit to prevent destroying earlier work.  These subcolumn collapses
produce the bulk of the zeroes and are done using {\bf $\clubsuit$Householder}.
After this, fewer than $d$ entries remain to be zeroed in the column
below the diagonal.  These are eliminated using a controlled
reflection containing $n-1$ controls and targeting the top line.
With the appropriate one-qudit Householder, the diagonal entry will
zero lower nonzero terms while the older zeroes in the column
are protected by the controls on the lower $n-1$ lines.

We now give a formal statement of the algorithm.
We emphasize the addition of controls when
previously generated circuits are incorporated into the universal circuit
(i.e. recursively telescoping control.)

\bigskip

\vbox{
\noindent
{\bf \hrule}

\noindent
Algorithm 4:  {\bf Triangle$(U,d,n)$}

\noindent
{\hrule}

\begin{tabular}{l}
{\bf if } $n=1$ {\bf then} \\
\quad Triangularize $U$ using a $QR$ reduction. \\
{\bf else} \\
\quad Reduce top-left $d^{n-1} \times d^{n-1}$ subblock using
{\bf Triangle$(\ast,d,n-1)$}, (writing output to bottom $n-1$ circuit lines) \\
\quad {\bf for} $k=0,1,\dots,d-1$ 
{\bf do} \quad \% {\em Block-column iteration} \\
\quad \quad {\bf for} columns $j=kd^{n-1},\dots,[(k+1)d^{n-1}-1]$ {\bf do} \\
\quad \quad \quad {\bf for} $\ell=(k+1),\dots,(d-1)$ 
{\bf do} \% {\em Block-row iterate} \\
\quad \quad \quad \quad Use $\clubsuit${\bf Householder} to zero the 
column entries $(k+\ell)d^{n-1},\dots,[(k+\ell+1)d^{n-1}-1]$, \\
\quad \quad \quad \quad \ \ {leaving a nonzero entry}
at $(k+\ell)c_2 \ldots c_n$ for $j=c_1c_2\ldots c_n$ and \\
\quad \quad \quad \quad \ \ {adding $\ket{k+\ell}$- control
on the most significant qudit}. \\
\quad \quad \quad {\bf end for} \\
\quad \quad \quad Clear the remaining nonzero entries below diagonal 
using one $\bigwedge[ Tc_2\ldots c_n, V]$. \\
\quad \quad {\bf end for} \quad 
\% {\em All subdiagonal entries zero in block-col} \\
\quad \quad Use {\bf Triangle}$(\ast,d,n-1)$ on the 
$d^{n-1}\times d^{n-1}$ matrix
at the $(k+1)^{\mbox{st}}$ block diagonal \\
\quad \quad \ \ {adding $\ket{k+1}$- control to the
most significant qudit}.
\\
\quad {\bf end for} \\
{\bf end if-else} \\
\end{tabular}

\noindent
{\hrule}
}
\bigskip

To generate a circuit for a unitary operator $U$, we use {\bf Triangle}
to reduce $U$ to
a diagonal operator $W=\sum_{j=0}^{d^n-1}
\mbox{e}^{i \theta_j} \ket{j} \bra{j}$.
Now $V$ and $U=WV$ would be indistinguishable if a von Neumann measurement
$\{ \ket{j}\bra{j} \}_{j=0}^{d^n-1}$ were made after each computation.
However, the diagonal is important if $U$ is a computation corresponding
to a subblock of the circuit of a larger computation with other trailing,
entangling interactions.  In this case, Figure \ref{fig:wedgek} makes
clear how to build a circuit for a controlled relative
phase $I_{d^n}+(\mbox{e}^{i\theta} - 1)\ket{j} \bra{j}$
in $O(n)$ gates.  Since $W$ has only $O(d^n)$ such phases, the 
corresponding circuit for $W$ costs $O(nd^n)$ two-qudit gates and as such
is asymptotically irrelevant to $\Theta(d^{2n})$.

\section{Counting Gates and Controls}
\label{sec:counts}

Let $h(n,k)$ be the number of $k$-controls required in the 
{\bf Single-$\clubsuit$Householder} 
reduction of some $\ket{\psi} \in \mathcal{H}_n$.  Then
clearly $h(n,k)=0$ for $k \geq 2$.  Moreover, each
$0$-control results from an element of the $\clubsuit$-sequence of
the form $00\ldots 0 \clubsuit \clubsuit \ldots \clubsuit$, and
there are $n$ such sequences.  Thus,
since the number of elements of the $\clubsuit$-sequence is
$(d^n-1)/(d-1)$, we see that
\begin{equation}
\left\{
\begin{array}{rcr}
h(n,1) & = & (d^n-1)/(d-1) - n \\
h(n,0) & = & n \\
\end{array}
\right.
\end{equation}

We next count controls in the matrix algorithm
{\bf Triangle} of \S \ref{sec:upper}.
We break the count into two pieces: $g$ for the work outside
the main diagonal blocks and $f$ for the work within.

Let $g(n,k)$ be the number of controls applied in 
operations in each column that zero the matrix below the
{block diagonal}; this is the total work in the {\bf for} $j$ loops
of {\bf Triangle}.  
We use
{\bf Single-$\clubsuit$Householder} $d(d-1)d^{n-1}/2$ times
since there are $d(d-1)/2$
blocks of size $d^{n-1} \times d^{n-1}$ below the block diagonal,
and we add a single control to those counted in $h$.
The last statement in the loop is executed $d^{n}-d^{n-1}$ times.
Therefore, letting
$\delta_j^k$ be the Kronecker delta, the counts are  
\begin{equation}
g(n,k) \ = \ \delta_k^{n-1} (d^{n}-d^{n-1})+
\frac{1}{2}d(d-1)d^{n-1} h(n-1,k-1) 
\end{equation}
Supposing $n \geq 3$, then we see that
\begin{equation}
\label{eq:subdiag}
g(n,k) \ = \ 
\left\{
\begin{array}{rr}
d^n-d^{n-1}, & k = n-1 \\
0, & n-1 \leq k \leq 3 \\
\frac{1}{2}d^n(d^{n-1}-1) - \frac{1}{2}d^n(d-1)(n-1), & k=2 \\
\frac{1}{2}d^n(d-1)(n-1), & k=1 \\
0, & k=0 \\
\end{array}
\right.
\end{equation}
Finally, let $f(n,k)$ be the total number of $k$-controlled operations
in the {\bf Triangle} reduction, including the block diagonals.  This
work includes that counted in $g$, plus a recursive call to
{\bf Triangle} before the {\bf for} $k$ loop, plus
$(d-1)$ calls within the $k$ loop, for a total of 
\begin{equation}
\label{eq:recurse}
f(n,k) \ = \ g(n,k) + f(n-1,k) + (d-1)f(n-1,k-1),
\end{equation}
with $f(n,0)=1$ and $f(1,k)=0$ for $n,k>0$.

Using the recursive relation of Equation \ref{eq:recurse} and the
counts of Equation \ref{eq:subdiag},
we next argue that {\bf Triangle} has no more than $O(d^{2n})$ controls.
Two lemmas are helpful.

\begin{lemma}
\label{lem:gross_overestimate}
For sufficiently large $n$, we have $f(n,k) \leq d^{2n-k+4}$.
\end{lemma}

\begin{proof}
By inspection of Equation \ref{eq:subdiag}, we see that
$g(n,k) \leq (1/2) d^{2n-k+2}$ for all $k$ and $n$ large.  
Now $f(n,0)=1$, which we take as an inductive hypothesis
while supposing $f(n-1,\ell)\leq d^{2n-2-\ell+4}=d^{2n-\ell+2}$.
Thus, using the recursion relation of Equation \ref{eq:recurse},
\begin{equation}
\begin{array}{lcl}
f(n,k) & \leq & \frac{1}{2} d^{2n-k+2} + d^{2n-k+2} + (d-1) d^{2n-k+3} \\
& = & d^{2n-k+4} \; 
\big( \; \frac{1}{2d^2} + \frac{1}{d^2} + 1 - \frac{1}{d} \; \big)
\\
\end{array}
\end{equation}
Now since $d> 3/2$, we must have $\frac{1}{d} > \frac{3}{2d^2}$, whence
an inductive proof of the result.
\end{proof}

\begin{lemma}
\label{lem:routine}
$\sum_{k=0}^{n-1} k f(n,k) \in O(d^{2n})$.
\end{lemma}

The proof of Lemma \ref{lem:routine} follows from
checking $\sum_{k=0}^{n-1} k d^{2n-k} \in O(d^{2n})$.  The latter
fact follows from either explicitly computing the sum by deriving the
appropriate geometric series or alternately using integral comparison.
Thus, the total number of control
boxes in the circuit digram grows as $O(d^{2n})$.  The theorem 
of the introduction asserting a size $O(d^{2n})$ universal
circuit composed of two-qudit gates follows, given the commentary of 
\S \ref{sec:uniform_control} on decomposing a $k$-controlled one-qudit operator
into two-qudit gates.

Figure \ref{fig:control_table} shows actual counts of
control boxes for specific instances of $d$, $n$.  These are illuminating
given that Lemma \ref{lem:gross_overestimate} overestimates the number
of $k$-controls for most $k$.  These counts are obtained using
a {\tt C++} implementation of the recursion presented in this section
and have been verified by an explicit {\tt MatLab} implementation of
the entire circuit synthesis algorithm for small $d$, $n$.

\begin{figure}[t]
{\scriptsize
\begin{tabular}{lr||r|r|r|r|r|r|r|r|r||}
\hline 
& $d$ & 2 & 3 & 4 & 5 & 6 & 7 & 8 & 9 & 10 \\
$n$ &   &   &   &   &   &   &   &   &   & \\
\hline
\hline
2 & & 5 & 17 & 39 & 74 & 125 & 195 & 287 & 404 & 549 \\
\hline 
3 & & 40 & 285 & 1 140 & 3 370 & 8 820 & 17 535 & 33 880 & 60 660 & 102 240 \\
\hline
4 & & 220 & 3 240 & 22 176 & 100 000 & 345 060 & 987 840 & 2 464 000 & 
5 528 736 & 11 407 500 \\
\hline
5 & & 1 040 & 32 130 & 379 776 & 2 631 500 & 12 931 920 & 49 999 110 & 
161 960 960 & 457 946 136 & \\
\hline
6 & & 4 560 & 301 239 & 6 220 032 & 66 768 750 & 470 221 200 & & & & \\
\hline
7 & & 19 200 & 2 757 807 & 100 279 728 & 1 676 043 750 & & & & & \\
\hline
8 & & 79 040 & 24 994 494 & 1 608 794 112 & & & & & & \\
\hline
9 & & 321 280 & 225 584 676 & & & & & & & \\
\hline
10 & & 1 296 640 & 2 032 525 629 & & & & & & & \\
\hline
11 & & 5 212 160 & 1 120 813 409 & & & & & & & \\
\hline
12 & & 20 904 960 & & & & & & & & \\
\hline
\hline
\end{tabular}
}
\caption{  
\label{fig:control_table}
Total number of control boxes in the new universal circuit
as a function of the level $d$ and number of qudits $n$
}
\end{figure}

\section{Conclusions}
\label{sec:conclusions}

We conclude with some remarks.
Locality in quantum mechanics is a function of the tensor (Kronecker)
product structure of the state space in question.  In quantum computing,
the Hilbert space factors are often finite dimensional.  Measuring
difficulty by counting two-particle interactions, we have generalized
a recent optimal asymptotic of $\Theta(2^{2n})$ for two-level quantum
bits to a new optimal asymptotic $\Theta(d^{2n})$ for $d$-level quantum
dits.  The result is exponentially better (asymptotically) 
than that obtained by emulating
such qudits with qubits, given $d \neq 2^\ell$.
This arises since the tensor decompositions are incompatible, except in
the case that $d$ is a power of $2$.

Multi-level quantum logics have been proposed as an alternative to qubits
due to the trade-off in the tensor structure.  For $d>2$, there is a larger
space of local operations, and fewer entangling gates might be required to
realize a target quantum computation (unitary evolution \cite{Hoyer:97}.)  
This work has
moreover demonstrated that such a benefit does not scale with the number
of particles $n$, but rather must consist (at most) of a constant
factor reduction in the number of required entangling gates.  However,
our result only applies to symmetry-less evolutions, and particular
computations might be better suited to certain multi-level and tensor
structures on Hilbert space than others.

\noindent 
{\bf Acknowledgements:}
We thank the
authors of {\tt quant-ph/0406003}, whose package {\tt Qcircuit.tex}
was used in creating this document's quantum circuit diagrams.
{DPO was supported in part
by the National Science Foundation under Grant CCR-0204084.  GKB was
supported in part by grant from ARDA/NSA.} 

\noindent
{\bf Implementations:}  {\tt MatLab} source code (``{\tt .m files}'')
have been included in the posting of this docoument to
{\tt http://www.arxiv.org}.  Please download the source format and
consult the {\tt README} file.  There is also a short {\tt C++}
program implementing the recursive control box counts.  These files
are very useful in understanding \S \ref{sec:upper}.

\noindent
{\bf NIST disclaimer. }
  Certain commercial equipment or instruments may be identified in this
  paper to specify experimental procedures.
  Such identification is not intended to imply recommendation
  or endorsement by the National Institute of Standards and
  Technology.

\appendix

\section{Proof of Correctness for State-Synthesis}
\label{app:correct}

We sketch the proof of correctness of the Algorithm for state-synthesis
employed to attain Equation \ref{eq:household_state}:
\begin{equation}
\prod_{k=1}^{p} \bigwedge[\mathcal{C}(p-k+1),V(p-k+1)] \; \ket{\psi}
\ = \ \ket{0}
\end{equation}
Given the Algorithm, $p=(d^n-1)/(d-1)$ is the number of elements
of the $\clubsuit$-sequence.  Suppose for clarity that $\ket{\psi}$ is
generic, so that no amplitudes (components) are zero at the outset.
Then it would suffice to prove (i) that each operator 
$\bigwedge[\mathcal{C}(j),V(j)]$
introduces $d-1$ new zeroes into the state $\ket{\psi_{j+1}}$ 
not present in $\ket{\psi_j}$ and (ii) 
$\bigwedge[\mathcal{C}(j),V(j)]$ does not act on 
previously zeroed entries.
The assertion (i) is straightforward and left to the reader.
However, the second assertion is \emph{false}.  Rather, the 
controlled one-qudit operators do act
on previously zeroed entries, but they
act in such a way that only linear combinations of these zeroes
are ever introduced as new amplitudes.
The discussion below makes this assertion precise and proves it.

To facilitate this, we label $S=\{0,1,\ldots,d^n-1\}$ the index set and
introduce the notation $S_\ast(j)$ for the set
of component indices of $\ket{\psi_j}$
that have not explicitly been reduced to a zero
by some $\bigwedge[\mathcal{C}(k),V(k)]$, 
$k \leq j$.  
We label $S[\mathcal{C}(j)]$ to be the set of control indices
of $\mathcal{C}(j)$, per Definition \ref{def:uniform_control}.
Also, define $\ell$ by $\mathcal{C}(j)_\ell = T$.
Now there is a group action of $\mathbb{Z}/d\mathbb{Z}$ on the index
set $S$ corresponding to addition mod $d$ on the $\ell^{\mbox{th}}$ dit:
\begin{equation}
c \; \bullet_\ell \ c_1 c_2 \ldots c_n \ = \ 
c_1 c_2 \ldots c_{\ell-1} (c_\ell + c \mbox{ mod d}) c_{\ell+1} \ldots c_n
\end{equation}
Since the operator $V(j)$ is applied to qudit $\ell$, the
amplitudes (components)
of $\ket{\psi_{j+1}}$ are either equal to the corresponding amplitude
of $\ket{\psi_j}$ or else are linear combinations of the
$\ket{\psi_j}$-amplitudes whose indices lie in the $\mathbb{Z}/d\mathbb{Z}$
orbit contained in $S[\mathcal{C}(j)]$.  Formally, we have 
proven the following Proposition.

\begin{proposition}
\label{prop:no_mixing}
Suppose 
\begin{equation}
\label{eq:action_sets}
\begin{array}{lcl}
(\mathbb{Z}/d\mathbb{Z}) \; \bullet_\ell \  \ 
S_\ast(j) \cap S[\mathcal{C}(j)] & \subseteq &
S_\ast(j) \cap S[\mathcal{C}(j)] \\
\end{array}
\end{equation}
(We remark that should theinclusion hold, then it is an equality.)  
Then $\ket{\psi_{j+1}}$ has at least $d-1$ more 
zero amplitudes than $\ket{\psi_j}$.
\end{proposition}

The final question is how one proves the appropriate set inclusions.
The point is to carefully understand the structure of
$S_\ast(j)$.   We will eventually prove that
$S_\ast(j)$ is the union of the three sets $R_1(j)$, $R_2(j)$,
and $R_3(j)$ below.  However, we define them independently,
as the induction technically requires the decomposition
at the $j^{\mbox{th}}$ step to avoid mixing as the next
operator is applied.

\begin{appdefinition}
Suppose the $j^{\mbox{th}}$ term of the $\clubsuit$-sequence is given
by $c_1 c_2 \ldots c_{\ell-1} \clubsuit \ldots \clubsuit$.
We have $\mathcal{C}(j)$ the corresponding control word,
with $\mathcal{C}(j)_\ell=T$.
Consider the following three sets, noting $R_1(j)$ may be vacuous.
\begin{equation}
\begin{array}{lcl}
R_1(j) & = & 
\bigsqcup_{q=0}^{\ell-2} \bigg\{ \ c_1 c_2 \ldots c_{q} k 0 0 \cdots 0 \; \;
; \; k < c_{q+1}, k \in \{0,1,\ldots,d-1\} \  \bigg\} \\
R_2(j) & = & 
\bigg\{ \ c_1 \cdots c_{\ell-1} k 0 0 \ldots 0 \; ; \;
k\in \{0,1,\ldots,d-1\} \ \bigg\} \\
R_3(j) & = & 
\bigg\{ \  f_1 \cdots f_{\ell-1} k_\ell k_{\ell+1} \ldots k_n \; ; \;
f_1 f_2 \ldots f_{\ell-1} > c_1 c_2 \cdots c_{\ell-1},
k_\ast \in \{0,1,\ldots,d-1\} \  \bigg\} \\
\end{array} 
\end{equation}
\end{appdefinition}

\begin{appremark}
These sets may be interpreted in terms of Figure
\ref{fig:Housegraph}.  Recall the figure recovers the $\clubsuit$-sequence
by doing a depth-first search of an appropriate tree.  In this context,
$S_\ast(j)$ is the set of nonzero components of $\ket{\psi_j}$ at the
$j^{\mbox{th}}$ node.  The subset $R_3(j)$ results from indices that
lie in nodes not yet traversed, loosely
below the present node in the tree or to the right.  
The set $R_2(j)$ is precisely
the set of indices in the current node, node $j$.  The set
$R_1(j)$ is the set of indices of elements that have
been previously used to zero other elements and still remain
nonzero themselves; 
it is the set of indices of elements
that were always at the top of nodes
already traversed in the depth-first
search.  Thus, $R_1(j)$ is loosely a set of entries within nodes to the
left and perhaps above node $j$.
\end{appremark}

\begin{lemma}
\label{lem:Sorbit}
Let $\mathcal{C}(j)$, $\ell$, be as above, and
label $\tilde{S}_\ast(j)=R_1(j) \sqcup R_2(j) \sqcup R_3(j)$.  Then
\begin{equation}
(\mathbb{Z}/d\mathbb{Z}) \; \bullet_\ell \  
\tilde{S}_\ast(j) \cap S[\mathcal{C}(j)]  \ \subseteq \  
\tilde{S}_\ast(j) \cap S[\mathcal{C}(j)]
\end{equation}
\end{lemma}

\begin{proof}
Due to the choice of a single control on
a dit to the right of position $\ell$ in the appropriate
term of the $\clubsuit$-sequence,
$R_1(j) \cap S[\mathcal{C}(j)]= \emptyset$.  On the other hand,
a direct computation verifies
that $(\mathbb{Z}/d\mathbb{Z})\bullet_\ell R_2(j) \subset
R_2(j)$ and also that
$R_2(j) \cap S[\mathcal{C}(j)]=R_2(j)$.

Finally, we note that
$(\mathbb{Z}/d\mathbb{Z})\bullet_\ell R_3(j) \subset R_3(j)$.
However, the following partition is in general nontrivial:
\begin{equation}
\label{eq:R3part}
R_3(j)\ = \ \{R_3(j) \cap S[\mathcal{C}(j)] \} \sqcup
\{R_3(j) \cap \big(S-S[\mathcal{C}(j)]\big) \}
\end{equation}
Should $\mathcal{C}(j)$ admit no control, we are done.  If not,
let $m<\ell$ be the control qudit,
i.e. $S[\mathcal{C}(j)]=\{m\}$.  Then
\begin{equation}
R_3(j) \cap S[\mathcal{C}(j)] \ = \ 
\bigg\{ \  f_1 \cdots f_{\ell-1} k_\ell k_{\ell+1} \ldots k_n \; ; \;
{\bf f_m = c_m},
f_1 f_2 \ldots f_{\ell-1} > c_1 c_2 \cdots c_{\ell-1},
k_\ast \in \{0,1,\ldots,d-1\} \  \bigg\}
\end{equation}
Hence the $\mathbb{Z}/d\mathbb{Z}$ action respects the partition
of Equation \ref{eq:R3part} as well.
\end{proof}

\vbox{
\begin{lemma}
\label{lem:Zlemma}
Let $\mathcal{C}(j)$, $\ell$, and $S_\ast(j)$ be as above,
with $\mathcal{C}(j)$ resulting from
$c_1 c_2 \ldots c_{\ell-1} \clubsuit \ldots \clubsuit \clubsuit$ of the
$\clubsuit$-sequence.
Let $\mathcal{Z}=
\{c_1 c_2 \ldots c_{\ell-1} k 0 0 \ldots 0 \; ; \; k\in \{1,2,\ldots,d-1\} \cap
\mathbb{Z} \}$ be the elements zeroed by $\bigwedge[\mathcal{C}(j),V(j)]$.
Then $R_1(j) \sqcup R_2(j) \sqcup R_3(j)=R_1(j+1) \sqcup R_2(j+1) \sqcup
R_3(j+1) \sqcup \mathcal{Z}$.
\end{lemma}
}

\begin{proof}
We break our argument into two cases based on the value of $c_{\ell-1}$.

\noindent
{\bf Case $c_{\ell-1} < d-1$:}
The $(j+1)^{\mbox{st}}$ term of the $\clubsuit$-sequence is
is given by \hbox{$c_1 c_2 \ldots (c_{\ell-1}+1) 0 0 \ldots 0 \clubsuit$}.
Note that for leaves of the tree, the buffering sequence of zeroes
is vacuous.
\begin{equation}
\begin{array}{lcl}
R_1(j+1) & = & R_1(j) \sqcup R_2(j) -\mathcal{Z} \\
R_2(j+1) \sqcup R_3(j+1) & = & R_3(j) \\
\end{array}
\end{equation}
 Hence
$R_1(j) \sqcup R_2(j)\sqcup R_3(j) = R_1(j+1) \sqcup R_2(j+1)
\sqcup R_3(j+1) \sqcup \mathcal{Z}$.

\noindent
{\bf Case $c_p=d-1$:}
Suppose instead the $j^{\mbox{th}}$ $\clubsuit$-sequence term is 
$c_1 c_2 \ldots c_{\ell-2} (d-1) \clubsuit \clubsuit \ldots \clubsuit$,
so that the $(j+1)^{\mbox{st}}$ term is
$c_1 c_2 \ldots c_{\ell-2} \clubsuit \clubsuit \clubsuit \ldots \clubsuit$.
We note that 
$\{c_0 c_1 \ldots c_{\ell-2} (d-1) 0 \ldots 0\} 
\in R_2(j) \cap R_2(j+1)$.\footnote{So in the application, the amplitude
(component) of this index is the single amplitude not zeroed by
$\bigwedge[\mathcal{C}(j),V(j)]$, but it is immediately afterwards
zeroed by $\bigwedge[\mathcal{C}(j+1),V(j+1)]$.}
Then
\begin{equation}
\begin{array}{lcl}
R_1(j) & = & R_1(j+1) \sqcup R_2(j+1) -\{ c_0 c_1 \ldots c_{\ell-2} (d-1)
0 \ldots 0\} \\
R_2(j) & = & \mathcal{Z} \sqcup \{c_0 c_1 \ldots c_{\ell-2} (d-1) 0 \ldots 0\} 
\\
R_3(j) & = & R_3(j+1) \\
\end{array}
\end{equation}
>From the first two, 
$R_1(j) \sqcup R_2(j) = R_1(j+1) \sqcup R_2(j+1) \sqcup \mathcal{Z}$.
Hence $R_1(j) \sqcup R_2(j)\sqcup R_3(j) = R_1(j+1) \sqcup R_2(j+1)
\sqcup R_3(j+1) \sqcup \mathcal{Z}$.
\end{proof}

\begin{proposition}
$S_\ast(j)= R_1(j) \sqcup R_2(j) \sqcup R_3(j)$ is the set of zero amplitudes
(components) of a generic $\ket{\psi_j}$.
\end{proposition}

\begin{proof}
The proof is by induction.  For $j=1$, we have
\begin{equation}
R_1(1)=\emptyset, \quad 
R_2(1)=\{00\ldots 0 \ast\}, \quad 
R_3(1)=\{ c_1 c_2 \ldots c_{n-1} \ast \; ; \; \mbox{ some } c_j > 0\}
\end{equation}
Hence the entire index set $S=S_\ast(1)= R_1(1) \sqcup R_2(1) \sqcup R_3(1)$.

Hence, we suppose by way of induction that
$S_\ast(j)=R_1(j) \sqcup R_2(j) \sqcup R_3(j)$ and attempt to
prove the similar statement for $j+1$.  
Now $\bigwedge[C(j),V(j)]$ will add new zeroes to the amplitudes
(components) with indices $\mathcal{Z}$ by Lemma \ref{lem:Zlemma}.
On the other hand, $\bigwedge[C(j),V(j)]$ will not destroy any
zero amplitudes existing in $S_\ast(j)$ 
due to the induction hypothesis,
Lemma \ref{lem:Sorbit}, and Proposition \ref{prop:no_mixing}.  Thus
$S_\ast(j+1)=R_1(j+1) \sqcup R_2(j+1) \sqcup R_3(j+1)$.
\end{proof}

\section{Unitary Circuits From State-Synthesis Circuits}
\label{sec:synth_to_uni}

In this appendix, we give an alternate construction
of optimal order circuits for unitary evolutions 
from
optimal circuits producing states per \S \ref{sec:Householder}.
We present a constructive procedure for building any
unitary $U \in \mathbb{C}^{d^n \times d^n}$ from $d^n$ copies of
an optimal state-synthesis circuit and other
asymptotically negligible subcircuits using an eigen-decomposition
of $U$ \cite{Knill_state}.
If the state-synthesis circuit
is any choice that contains $O(d^n)$ gates, then the resulting $O(d^{2n})$
circuit for $U$ is optimal.

Let 
$\{ \lambda_j = \mbox{e}^{i \theta_j} \}_{j=0}^{d^n-1}$
be the eigenvalues of $U$, with $\{ \ket{\lambda_j} \}_{j=0}^{d^n-1}$ 
a corresponding set of orthonormal eigenvectors.  
We suppose circuits containing
$O(d^n)$ two-qudit gates for unitary operators
$W_j$ with $W_j \ket{j} = \ket{\lambda_j}$, $0 \leq j \leq d^n-1$.
Then for a second set of phasing unitaries $P_j$, we may write
\begin{equation}
\begin{array}{lcl}
W_j & = & \ket{\lambda_j} \bra{j} + \sum_{k \neq j} \ket{\psi_{j,k}}
\bra{k} \\
P_j & = & I_{d^n-1} + 
(\mbox{e}^{i \theta_j} - 1) \ket{j} \bra{j} \\
\end{array}
\end{equation}
Then by unitarity, $\bra{\psi_{j,k}} \lambda_j\rangle= 0$ for all
$j,k$.  Now note that
$W_j P_j W_j^\dagger=\mbox{e}^{i \theta_j} \ket{\lambda_j} \bra{\lambda_j}
+ \sum_{k \neq j} \ket{\psi_{j,k}}\bra{\psi_{j,k}}=
\mbox{e}^{i \theta_j} \ket{\lambda_j} \bra{\lambda_j}
+ \sum_{k \neq j} \ket{\lambda_j}\bra{\lambda_j}$.  Similarly by
induction, the following equality may be verified:
\begin{equation}
(W_0 P_0 W_0^\dagger) (W_1 P_1 W_1^\dagger) \ldots
(W_k P_k W_k^\dagger) \ = \ 
\sum_{j=0}^k \mbox{e}^{i \theta_j} \ket{\lambda_j} \bra{\lambda_j}
+ \sum_{j=k+1}^{d^n-1} \ket{\lambda_j} \bra{\lambda_j}
\end{equation}
Then considering the eigendecomposition of $U$ and taking
$k=d^n-1$, we have the following factorization:
\begin{equation}
\label{eq:spectraldecomposition}
U \ = \ \prod_{j=0}^{d^n-1} W_j P_j W_j^\dagger
\end{equation}
Now note that the techniques of \S \ref{sec:uniform_control} allow
for realization of $P_j$ in $O(n)$ two-qudit gates.  Thus, since
by hypothesis each $W_j$ admits a size $O(d^n)$ circuit, the
circuit corresponding to Equation \ref{eq:spectraldecomposition}
contains $O(d^{2n})$ two-qudit gates and is asymptotically optimal.

As a remark, the circuit synthesis procedure might take
$\ket{0} \mapsto \ket{\lambda_j}$ rather than 
$\ket{j} \mapsto \ket{\lambda_j}$.  However an $O(d^n)$ circuit for the
latter extracted from an $O(d^n)$ circuit for the former follows
from the similarity transform by a local
unitary per \S \ref{sec:Householder}.

\section{Optimal Asymptotics for Qudit Chains}
\label{sec:architectures}

The optimal asymptotic of $\Theta(d^{2n})$ also holds for more
restrictive gate libraries reflecting a choice of architecture.  We
note this in passing, focusing on the qudit chain architecture.

In the interest of being brief, we do not use formal definitions.
Note that the body shows that the library $\mbox{LU} \sqcup 
\{ \bigwedge_1( \sigma^x \oplus I_{d-2})\}$ is qudit universal, where
$\mbox{LU} = \otimes_1^n SU(d)$ and we intend any instantiation of
$\bigwedge_1(\sigma^x \oplus I_{d-2})$ to be allowed.  An
\emph{architecture} will here refer to a restriction on this gate
library.  In particular, one supposes that the qudits correspond
to the vertices of some graph, which by hypothesis is connected.  Then only
the instantiations of $\bigwedge_1(\sigma^x \oplus I_{d-2})$ which
correspond to edges of the graph are allowed.  Since we may construct
qudit {\tt SWAP} between qudits connected by an edge, the restricted
library is also qudit-universal.  However, the asymptotics of the
library gates might be different from the asymptotics of the
standard gates.  Loosely, instantiations of
$\bigwedge_1(\sigma^x\oplus I_{d-2})$ between qudits $O(n)$ vertices
apart will now cost $O(n)$ gates rather than one, since
$O(n)$ {\tt SWAP}s between adjacent qudits are also needed.

The notion of a sub-architecture follows by comparing graphs and
subgraphs.  Thus, if a qudit chain is the architecture of a linear
sequence of qudits with consecutive qudits joined by edges, then the
qudit chain is a subarchitecture of a finite square, hexagonal, or
cubic lattice.  If a sub-architecture contains every vertex, then
asymptotics of the smaller architecture are at least as good as those
of the larger.  For the inclusion only admits more possible two-qudit
gates.

Thus consider in particular a
\emph{qudit chain.}  Suppose further the ordering of
the dits implicit in earlier notations, e.g. $d_1 d_2 d_3 \ldots d_n$,
is now referring to the architecture as well.  Thus given the
architectural restriction, using {\tt SWAP}s we see that an instantiation
of $\bigwedge_1(\sigma^x \oplus I_{d-2})$ costs $O(|j-k|)$
architecture gates if controlled on qudit $j$ and targeting qudit $k$,
rather than the old count of one gate.  A similar comment applies to
any two-qudit gate acting between qudits $j$, $k$.

By Appendix \ref{sec:synth_to_uni}, the $\Theta(d^{2n})$
asymptotic follows if we show that the
$\clubsuit$-Householder reduction requires only $O(d^n)$ 
architecture-local two-qudit gates.  Hence let $a(d,n,k)$ denote the number of
length $k$ singly-controlled
$\bigwedge_1(V)$ specified by the club sequence, where a two-qudit operator
acting on qudits $j$, $\ell$, has length $j+\ell+1$.  
As an example, the operator of Figure \ref{fig:club_to_control} is length four.
Now for most $k$, a length $k$ operation within the
$(n+1)^{\mbox{st}}$ club sequence results from a sequence of $k-1$ zeroes
in some term of the $n^{\mbox{th}}$ sequence in one of two ways: 
\begin{itemize}
\item  A length $n$ term of the form $d_2 d_3 \ldots d_j 0 0 \ldots 0 \clubsuit
\ldots \clubsuit$ is preprended to become
$d_1 d_2 \ldots d_j 0 0 \ldots 0 \clubsuit \ldots \clubsuit$.
\item  The length $n$ term of the form $00 \ldots 0 \clubsuit \ldots \clubsuit$
is prepended to become $d_1 0 0 \ldots 0 \clubsuit \ldots \clubsuit$.
Here, $d_1 \neq 0$.
\end{itemize}
Noting this structure, we produce the following recursion relations,
which completely determine $a(d,n,k)$:
\begin{equation}
\left\{
\begin{array}{lcl}
a(d,n+1,k) & = & (d-1) + d \; a (d,n,k) \\
a(d,n,0) & = & n \\
a(d,n,k) & = & 0 \mbox{ if } k \geq n-1 \\
\end{array}
\right.
\end{equation}
Now $a(d,n,0)=n$ does not factor into the recursive structure of the
other $a(d,n,k)$.  Rather, evaluating the recursion explicitly for 
$1 \leq k \leq n-1$, we obtain
\begin{equation}
a(d,n,k) \ = \ (d-1) \sum_{\ell=0}^{n-k-1} d^\ell \ = \ d^{n-k}-1
\end{equation}
We finally use this recursion to obtain our main result.

Indeed, note that since a length $k$ singly-controlled operation
may be realized in $O(k)$ local gates, it suffices to prove that
$\sum_{\ell=0}^{n-1} k \; a(d,n,k) = \sum_{\ell=0}^{n-1} k (d^{n-k}-1)$
is a function within $O(d^n)$.  This follows by either deriving the
appropriate geometric series in order to obtain the exact sum 
$\sum_{\ell=0}^{n-1} k d^{n-k}$ or
alternately by integral comparison of this second sum.  Thus, 
even in the chain gate
library, the $\clubsuit$-Householder reduction requires no more than
$O(d^n)$ gates and hence recovers an optimal 
state-synthesis asymptotic of $\Theta(d^n)$.  Consequently,
Appendix \ref{sec:synth_to_uni} produces an asymptotic of
$\Theta(d^{2n})$ chain architecture-restricted gates for any unitary
evolution $U \in U(d^{2n})$.


\begin{thebibliography}{99}

\bibitem{Deutsch:89}
D. Deutsch, 
Quantum Computational Networks,
{\em Proc. R. Soc. London A} {\bf 425}, 73 (1989).

\bibitem{Feynman:86}
R. P. Feynman,
The Computer as a Physical System:  a microscopic quantum-mechanical
Hamiltonian model of computers as represented by Turing-machines,
{\em Found. Phys.} {\bf 16}, 507 (1986), \\
P. Benioff, 
{\em J. Stat. Phys.} {\bf 22}, 563 (1980).

\bibitem{Gottesman:99}
D. Gottesman, 
Fault-Tolerant Quantum Computation with Higher-Dimensional Systems,
{\em Chaos, Solitons Fractals} {\bf 10}, 1749 (1999).
{\tt http://www.arxiv.org/abs/quant-ph/9802007 }

\bibitem{Blume-KahoutEtAl:02}
R. Blume-Kahout, C.M. Caves, I.H. Deutsch, 
Climbing Mount Scalable: Physical Resource Requirements for a 
Scalable Quantum Computer,
{\em Found. Phys.} {\bf 32}, 1641 (2002).

\bibitem{Lawler:64}
E.L. Lawler, Jour. ACM {\bf 11}, 431 (1964).\\
M. Fujita, Y. Matsunaga, and M. Cieseilski,
Mulit-Level Logic Optimization, Chap. 2 of
\emph{Logic Synthesis and Verification,} edited
by Soha Hassoun and Tsutomu Sasao,
Kluwer Academic Press, Norwell Massachusettes,
2001.

\bibitem{Brylinski:02}
J.-L. Brylinski and R.~Brylinski, Mathematics of Quantum Computation, edited by R.~Brylinski and G.~Chen, CRC Press (2002), {\tt http://www.arxiv.org/abs/ quant-ph/0108062}.

\bibitem{DiVincenzo:95}
D.P. DiVincenzo, 
Two-qubit Gates are Universal for Quantum Computation,
{\em Phys. Rev. A} {\bf 51}, 1015 (1995).

\bibitem{NielsenChuang:00}
      	M. Nielsen and I. Chuang,
      	{\em Quantum Computation and Quantum Information},
      	(Cambridge Univ. Press, 2000.)

\bibitem{Hoyer:97}
P. Hoyer, 
Efficient Quantum Transforms,
{\tt http://www.arxiv.org/abs/quant-ph/9702028} 

\bibitem{Knill:96}
E. Knill,
Non-binary Unitary Error Bases and Quantum Codes,
{\tt http://www.arxiv.org/abs/quant-ph/9608048}

\bibitem{Vartiainen:04}
J.J.Vartiainen, M.M{\"o}tt{\"o}nen, M.M.Salomaa,
Efficient Decomposition of Quantum Gates
{\em Phys. Rev. Lett.} {\bf 92}, 17902 (2004).

\bibitem{MuthukrishnanStroud:00}
A. Muthukrishnan and C.R.Stroud Jr.,
Multivalued Logic Gates for Quantum Computation
{\em Phys. Rev. A} {\bf 62}, 052309 (2000).

\bibitem{ShendeMarkovBullock:03}
V.V. Shende, I.L. Markov, S.S. Bullock,
Minimal Universal Two-qubit Controlled-not Based Circuits, 
{\em Phys. Rev. A} {\bf 69} 062321 (2004).
{\tt quant-ph/0308033}

\bibitem{BullockMarkov:04}
S.S.Bullock and I.L. Markov,
Asymptotically Optimal Circuits for Arbitrary n-qubit Computations,
{\em Quant. Inf. and Comp.} {\bf 4} 27 (2004).

\bibitem{BarencoEtAl:95}
A. Barenco, C. Bennett, R. Cleve, D. DiVincenzo, N. Margolus, P. Shor,
T. Sleator, J. Smolin, and H. Weinfurter,
Elementary Gates for Quantum Computation,
{\em Phys. Rev A} {\bf 52} 3457 (1995).

\bibitem{Cybenko:01}
G. Cybenko, 
Reducing Quantum Computations to Elementary Unitary Operations,
{\em Comp. in Sci. and Eng.} {\bf 27} March/April 2001.

\bibitem{BrennenOLearyBullock:04}
G.K.Brennen, D.P.O'Leary, S.S.Bullock, 
Criteria for Exact Qudit Universality,
{\tt http://www.arxiv.org/abs/quant-ph/0407223} 

\bibitem{SchirmerEtAl:03}
S.G. Schirmer, A.D. Greentree, and D.K.L.Oi,
Implementation of Controlled Multi-qudit Operations for a Solid-state 
Quantum Computer Based on Charge Qudits,
{\tt http://www.arxiv.org/abs/quant-ph/0305052}.

\bibitem{ShapiroEtAl:03}
E. A. Shapiro, I. Khavkine, M. Spanner, and M. Yu. Ivanov,
Strong-field Molecular Alignment for Quantum Logic and Quantum Control,
{\em Phys. Rev. A} {\bf 67} 013406 (2003).

\bibitem{BartlettEtAl:02}
S.D. Bartlett, H. de Guise, and B.C. Sanders,
Quantum Encodings in Spin Systems and Harmonic Oscillators,
{\em Phys. Rev. A} {\bf 65} 052316 (2002).

\bibitem{KloseEtAl:01}
G. Klose, G. Smith, and P. S. Jessen,
Measuring the Quantum State of a Large Angular Momentum,
{\em Phys. Rev. Lett.} {\bf 86} 4721 (2001).

\bibitem{Aharonov:01}
D.~Aharonov, Presented at {\em Conference on Quantum Information: 
Entanglement, Decoherence and Chaos}, Institute for Theoretical Physics, 
Santa Barbara, 2001 (unpublished).

\bibitem{gvl}
G.H. Golub and C. van Loan, 
\emph{Matrix Computations}, Johns Hopkins Press,
1989.

\bibitem{Mottonen:04}
M. M{\"o}tt{\"o}nen, J. J. Vartiainen, V. Bergholm, M. M. Salomaa,
Quantum circuits for general multiqubit gates, \\
{\em Phys. Rev. Lett.} {\bf 93} 130502 (2004).
{\tt http://www.arxiv.org/abs/quant-ph/0404089}.

\bibitem{Mottonenb:04}
M. M{\"o}tt{\"o}nen, J. J. Vartiainen, V. Bergholm, M. M. Salomaa,  \\
Transformation of quantum states using uniformly controlled rotations, \\
{\tt http://www.arxiv.org/abs/quant-ph/0407010} 

\bibitem{ShendeBullockMarkov:04}
V.V. Shende, S.S. Bullock, I.L. Markov,
A Practical Top-down Approach to Quantum Circuit Synthesis, \\
{\tt http://www.arxiv.org/abs/quant-ph/0406176}

\bibitem{Knill_state}
E. Knill,
Approximation by Quantum Circuits,
{\tt http://www.arxiv.org/abs/quant-ph/9508006}

\bibitem{Deutsch_state}
D. Deutsch, A. Barenco, A. Ekert,
Universality in Quantum Computation,
{\em Proc. R. Soc. London A} {\bf 449}, 669 (1995).


\end{thebibliography}
\end{document}